# On the Structure of Vacancy Ordered Superconducting Alkali Metal Iron Selenide

P. Zavalij[1], Wei Bao[2,*], X. F. Wang[3], J. J. Ying[3], X. H. Chen[3], D. M. Wang[2], J. B. He[2], X. Q. Wang[2], G.F Chen[2], P-Y Hsieh[4], Q. Huang[5] and M. A. Green[4,5,*]

[1] Department of Chemistry, University of Maryland, College Park, MD 20742 USA
[2] Department of Physics, Renmin University of China, Beijing 100872, China
[3] Hefei National Laboratory for Physical Science at Microscale and Department of Physics, University of Science and Technology of China, Hefei, Anhui 230026, China
[4] Department of Materials Science and Engineering, University of Maryland, College Park, MD 20742 USA
[5] NIST Center for Neutron Research, NIST, 100 Bureau Dr., Gaithersburg, MD 20878 USA

With single crystal X-ray diffraction studies, we compare the structures of three sample showing optimal superconductivity, $K_{0.774(4)}Fe_{1.613(2)}Se_2$, $K_{0.738(6)}Fe_{1.631(3)}Se_2$ and $Cs_{0.748(2)}Fe_{1.626(1)}Se_2$. All have an almost identical ordered vacancy structure with a ($\sqrt{5}$ x $\sqrt{5}$ x 1) super cell. The tetragonal unit cell, space group I4/m, possesses lattice parameters at 250K of $a = b = 8.729(2)$ Å and $c = 14.120(3)$ Å, $a = b = 8.7186(12)$ Å and $c = 14.0853(19)$ Å and at 295 K, $a = b = 8.8617(16)$ Å and $c = 15.304(3)$ Å for the three crystals, respectively. The structure contains two iron sites; one is almost completely empty, whilst the other is fully occupied. There are similarly two alkali metal sites that are occupied in the range of 72.2(2) % to 85.3(3) %. The inclusion of alkali metals and the presence of vacancies within the structure allows for considerable relaxation of the $FeSe_4$ tetrahedron, compared with members of the Fe(Te, Se, S) series, and the resulting shift of the Se – F – Se bond angles to less distorted geometry could be important in understanding the associated increase in the superconducting transition temperature. The structure of these superconductors distinguishes themselves from the structure of the non-superconducting phases by an almost complete absence of Fe on the (0 0.5 0.25) site as well as lower alkali metal occupancy that ensures an exact $Fe^{2+}$ oxidation state, which are clearly critical parameters in the promotion of superconductivity.

A new series of superconductors, with the general formula, $A_xFe_ySe_2$, has recently been identified and shown to have significantly enhanced superconducting transition temperatures ($T_c$ up to 32 K)[1] as compared to FeSe ($T_c \approx 8.5$ K)[2]. This series joins a growing list of pnictide and chalcogenides superconductors that include the original ZrCuSiAs-type compounds, such as $LnO_{1-x}F_xFeAs$[3-9], where Ln is a lanthanide ion, the $ThCr_2Si_2$-type structures of $AFe_2As_2$, where A is a alkali or alkaline earth metal[10-12] and the anti-PbFCl structure of AFeAs, where A is an alkali metal[13, 14]. All of these compounds contain iron in tetrahedral coordination that are edge-shared to form two-dimensional structures. The simplest of the chalcogenides are those with the anti-PbO structure with the general formula, $Fe_{1+x}$(Te, Se, S). FeSe is superconducting at 8.5 K[2] and is very sensitive to interstitial iron ions[15]. $Fe_{1+x}Te$, where x lies between 0.076 and 0.18, shows commensurate or incommensurate antiferromagnetism depending on exact composition[16]. The antiferromagnetism can be suppressed and superconductivity emerges on substituting some Te with either Se[17, 18] or S[19]. In both cases, the chemical pressure afforded by the Se[20] and S[21] inclusion reduces the amount of interstitial iron. The amount of the interstitial iron can be reduced by topotactic deintercalation using iodine[22]. $Fe_{1+x}Te_{0.7}Se_{0.3}$ was shown to transform from gapless paramagnetism to superconductivity on Fe removal[23].

The superconducting K-Fe-Se phases were found to possess the $ThCr_2Si_2$ structure and isostructural with $AFe_2As_2$[1]. The first structure determination of AFeSe in this layered tetragonal arrangement was with A as the group 13 metal, Tl. It possesses lattice parameters of $a = b = 3.890(1)$ Å and $c = 14.00(1)$ Å and was shown to be a p-type Pauli paramagnetic metal[24-26]. A related composition was shown to be antiferromagnetic at $T_N \approx 450$ K and have ordered Fe vacancies, which is best described as $TlFe_{2-x}Se_2$ with a $\sqrt{5}$ x $\sqrt{5}$ x 1 supercell giving a 5 fold volume increase at x ≈ 0.3[27]. Superconducting potassium iron selenide was first reported above 30 K in $K_xFe_2Se_2$, which has K deficiency, but stoichiometric Fe and Se, and lattice parameters of $a = 3.9136(1)$ Å and $c = 14.0367(7)$ Å[1]. Similar properties have also been reported in iron deficient $(Tl,K)Fe_xSe_2$[28], alkali metal deficient $Rb_{0.8}Fe_2Se_2$[29], $K_xFe_2Se_2$[30], and $K_{0.8}Fe_{2.3}Se_2$[31], and alkali metal and Se deficiency $Rb_{0.78}Fe_2Se_{1.78}$[32], and $Cs_{0.8}(FeSe_{0.98})_2$[33].

As a result of both the critical importance of interstitial iron in the $Fe_{1+x}$(Te, Se, S) series and the variation of properties in the $A_xFe_ySe_2$ series warrants careful determination of the exact stoichiometry of these superconductors. In a single crystal X-ray diffraction structure determination, we observe that in superconducting samples of composition, $K_{0.774(4)}Fe_{1.613(2)}Se_2$, $K_{0.738(6)}Fe_{1.631(3)}Se_2$ and $Cs_{0.748(2)}Fe_{1.626(1)}Se_2$, with a near perfect diamagnetic response[34], an ordered vacancy structure that contains both alkali metal and iron deficiencies to form an ordered ($\sqrt{5}$ x $\sqrt{5}$ x 1) super lattice, which is consistent with a recent transmission electron microscope study that reports a superstructure modulation along the [310] direction[35], as well as powder neutron diffraction[36]. The structure is related to a recent X-ray diffraction study of non-superconducting $K_{0.8+x}Fe_{1.6-y}Se_2$, but here we show that the superconducting composition contain one occupied and one vacant site, rather than two partially occupied Fe sites[37]. The complexity of the structure demonstrates the need to map out the iron selenide

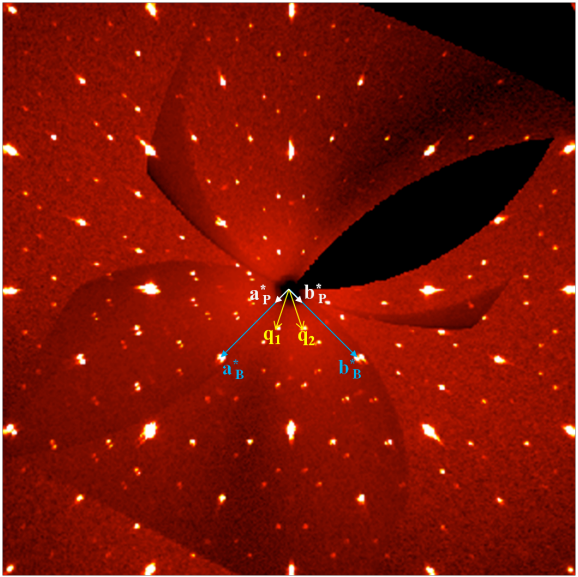 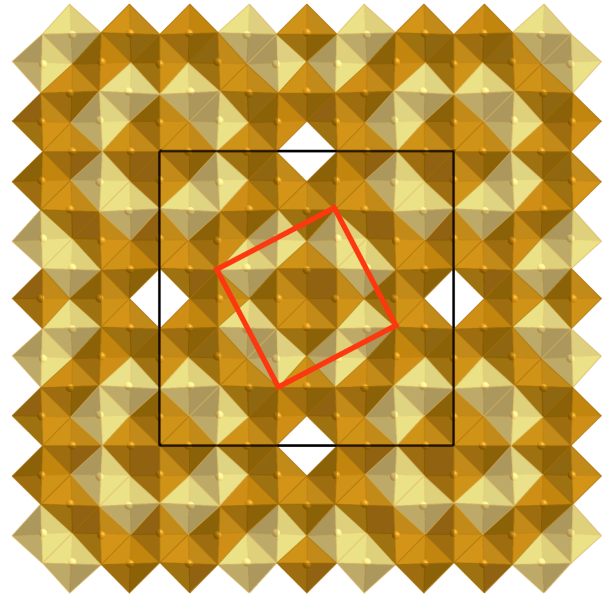

**Figure 1** Panel (a) Single Crystal X-ray Diffraction pattern of $K_{0.774(4)}Fe_{1.613(2)}Se_2$ showing the relationship between the original $ThCr_2Si_2$ cell of approximate dimensions 3.9 Å and 14 Å indicated by vectors $a_B^*$ and $b_B^*$ can be interpreted in (5 x 5 x 1) cell labeled, $a_p^*$ and $b_p^*$. Alternatively $q_1$ and $q_2$ represent two alternative twinned cell with ($\sqrt{5}$ x $\sqrt{5}$ x 1) as proposed by Haggstrom et al[27]. Panel (b) Diagramatic representation of the structure in real space showing the (5 x 5 x 1) cell (black) compared with the reduced ($\sqrt{5}$ x $\sqrt{5}$ x 1) cell (red).

phase diagram with great precision to evaluate the structure – property relations.

The single crystals used in this study were synthesized by first reacting Fe and Se to form FeSe, which were subsequently reacted with K or Cs in a similar manner to that previously reported. Previous papers have reported superconductivity for the crystals, $K_{0.774(4)}Fe_{1.613(2)}Se_2$[38], $K_{0.738(6)}Fe_{1.631(3)}Se_2$[31] and $Cs_{0.748(2)}Fe_{1.626(1)}Se_2$[38]. The X-ray intensity data at 250 K of metallic black plate-like crystal of $K_{0.774(4)}Fe_{1.613(2)}Se_2$ and $K_{0.738(6)}Fe_{1.631(3)}Se_2$ and at 295 K for $Cs_{0.748(2)}Fe_{1.626(1)}Se_2$, were measured on a Bruker Smart Apex2, CCD system equipped with a graphite monochromator and a $MoK_\alpha$ fine focus sealed tube ($\lambda = 0.71073$ Å). The exact details of the refinements are given in crystal structure reports in supplementary information.

The initial diffraction pattern for the very similar potassium and cesium iron selenide crystals, shown in Figure 1, could be indexed on a 25 times two dimensional modulation of the $ThCr_2Si_2$ structure. Initially crystal structure determination was performed in the super cell within space group I4/mmm. The latter was selected based on the intensity statistics. The resulting structure shown in Figure 1, panel b contains 6 Fe sites; 4 are fully occupied, one is fully vacant whilst one is partially occupied for about 40%. The final R-factor was 3.3% for the K containing structure. However, these results were inconsistent with the initial model proposed by Haggstrom et al for the related $TlFe_{2-x}Se_x$ structure[27], recent NMR work that suggested limited iron site disorder from partial occupancy[39], and powder neutron diffraction data[36]. In particular, the powder diffraction data, which can very accurately determine Fe composition in these systems[22, 23], refined in the I4/m space group, using the original ($\sqrt{5}$ x $\sqrt{5}$ x 1) super lattice. In addition, irreducible representation analysis of the magnetic structure in the 25 times cell suggested a reduced *ab* moment in two of the Fe sites, that was unphysical and warranted more detailed investigation. Further analysis of the crystal structure in the light of other measurements, showed that the structure can be represented in the ($\sqrt{5}$ x $\sqrt{5}$ x 1) cell with the lower I4/m space group, with the presence of twinning (180° rotation around 100 axis) with approximately 1:1 ratio of twin components. This structure then yields an ordered distribution of the two Fe sites, such that one is almost completely empty, whereas one is fully occupied. The relationship between the two cells is shown in Figure 1, panel b.

On the final stages of refinement of both K and Cs structures the difference Fourier maps showed presence of electron density in the almost empty Fe sites on the level of 1-2 electrons. When placing Fe atom in this site with all occupation factors including factors of K or Cs refined independently, the resulting composition satisfied charge balance very accurately (within one standard deviation). Thus the general formula could be written as $A_xFe_{1-x/2}Se_2$ (A = K, Cs) and in the final refinement the occupation of Fe and M were constrained to match this composition exactly. The final R-factor was 1.9 %, 3.38 %, and 3.02 % for $K_{0.774(4)}Fe_{1.613(2)}Se_2$, $K_{0.738(6)}Fe_{1.631(3)}Se_2$ and $Cs_{0.748(2)}Fe_{1.626(1)}Se_2$ crystals, respectively.

A diagrammatic representation of the structure of these alkali metal iron selenides that possesses similar body centered $ThCr_2Si_2$ stacking to $BaFe_2As_2$, are compared with the primitive symmetry of anti-PbFCl structure of NaFeAs, in Figure 2. These structures have the same morphology as $BaFe_2As_2$. However, the 5 times larger cell now has two crystallographic inequivalent positions for potassium, iron and selenium. Of these two iron positions, one Fe sites at (0 0.5 0.25) is almost completely empty for all three superconducting structures, while the other on a general position is fully occupied. A list of the atomic coordinates is given in Table 1.

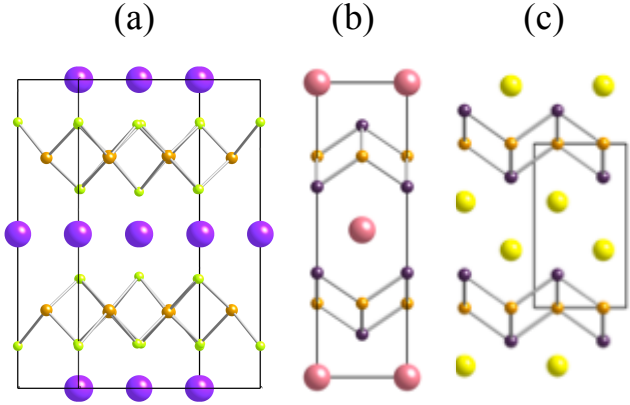

**Figure 2** Comparison of the (a) $A_xFe_{1-x/2}Se_2$ structure with the same body centered $ThCr_2Si_2$-type stacking as (b) $BaFe_2As_2$ rather than the primitive anti-PbFCl structure of (c) NaFeAs.

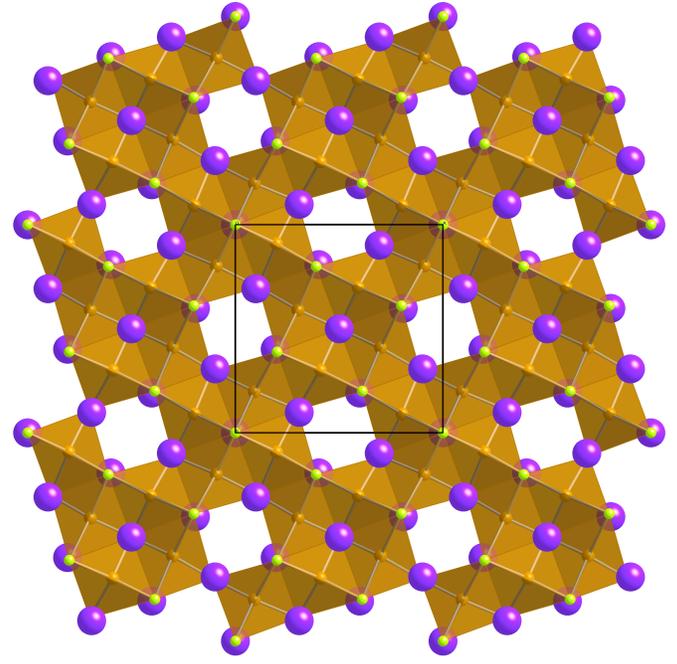

**Figure 3** Structure of $K_{0.774(4)}Fe_{1.613(2)}Se_2$ from the [001] direction in the ($\sqrt{5}$ x $\sqrt{5}$ x 1) cell showing fully occupied Fe sites (orange) decorated with ordered vacancy sites.

|  | $K_{0.738(6)}Fe_{1.613(2)}Se_2$ | $K_{0.774(4)}Fe_{1.613(2)}Se_2$ | $Cs_{0.748(2)}Fe_{1.626(1)}Se_2$ |
|---|---|---|---|
| A1 U | 0.0343(11) | 0.0337(6) | 0.0402(4) |
| A1 occ | 0.790(4) | 0.838(4) | 0.853(3) |
| A2 x | 0.8082(3) | 0.80817(17) | 0.80787(11) |
| A2 y | 0.4021(3) | 0.40207(15) | 0.40553(9) |
| A2 U | 0.0317(7) | 0.0292(4) | 0.0370(2) |
| A2 occ | 0.724(7) | 0.759(4) | 0.722(2) |
| Fe1 U | 0.017(6) | 0.014(7) | 0.007(4) |
| Fe1 occ | 0.078(7) | 0.032(4) | 0.065(2) |
| Fe2 x | 0.30149(8) | 0.30160(4) | 0.30160(8) |
| Fe2 y | 0.40856(11) | 0.40900(6) | 0.40855(9) |
| Fe2 z | 0.24696(7) | 0.24674(3) | 0.24791(5) |
| Fe2 U | 0.0178(2) | 0.01449(11) | 0.02017(17) |
| Se1 z | 0.13787(10) | 0.13743(5) | 0.14791(8) |
| Se1 U | 0.0177(2) | 0.01447(12) | 0.0199(2) |
| Se2 x | 0.11074(12) | 0.11118(7) | 0.11022(8) |
| Se2 y | 0.29968(6) | 0.29972(3) | 0.29798(6) |
| Se2 z | 0.35480(4) | 0.35469(2) | 0.34434(3) |
| Se2 U | 0.01922(15) | 0.01623(8) | 0.02176(13) |
| a (Å) | 8.729(2) | 8.7186(12) | 8.8617(16) |
| c (Å) | 14.120(3) | 14.0853(19) | 15.304(3) |
| T (K) | 250(2) | 250(2) | 295(2) |
| Se – Fe – Se bond angles | 107.79(5)° | 108.06(3)° | 109.28(4)° |
|  | 113.62(4)° | 113.62(2)° | 112.98(4)° |
|  | 113.69(3)° | 113.76(2)° | 112.60(3)° |
|  | 106.89(6)° | 106.60(3)° | 105.66(4)° |
|  | 113.71(4)° | 113.774(18)° | 112.69(3)° |
|  | 100.97(6)° | 100.79(3)° | 103.37(4)° |

**Table 1** Summary of atomic coordinates some selected bond angles in the tetragonal I4/m cell for the superconducting phases, $K_{0.774(4)}Fe_{1.613(2)}Se_2$, $K_{0.738(6)}Fe_{1.631(3)}Se_2$ and $Cs_{0.748(2)}Fe_{1.626(1)}Se_2$. A represents either K or Cs. Atomic positions are at A1 (0 0 5), A2 (x, y, 0.5), Fe1 (0 0.5 0.25), Fe2 (x ,y, z), Se1 (0 0 z) and Se2 (x y z). Both A positions are partially occupied. Fe1 is occupied with very low amounts, whereas Fe2 has full occupancy in all refinements

The Se – Fe – Se bond angles have extensive distribution ranging from 100.79° to 113.76°, but with a number of bond angles remarkable close to the ideal tetrahedron angle of 109.47°. This far less distorted arrangement than in the related Fe(Te, Se, S) series may offer an explanation to the considerably higher superconducting transition temperature; the sensitivity of the superconducting transition temperature with the iron anion tetrahedron has been suggested for the iron based superconductors[40-42].

The two iron sites in both the K and Cs superconducting crystals have very different occupancies. The general Fe *16i* site is fully occupied, whilst there is very low occupancy of the *4d* site, which is in contrast to that observed for non-superconducting samples.[37] Although the refinement of the *4d* Fe occupancy made a notably improvement to the fit, such low levels of iron (3.2 – 7.8%) may not be a bulk effect or homogeneously distributed within the crystal. Such occupancies could, for example, arise from small domains, such as near the crystal surface, which support the occupancy of the *4d* site, leaving the rest of the crystal without *4d* iron. In either the case of the presence of domains or where the crystal possesses a homogeneous distribution of *4d* iron, this arrangement is grossly different to that observed for non-superconducting samples in reference 37. Two significant differences can highlighted between the superconducting and non-superconducting compositions. Firstly, the non-superconducting samples contain significantly more alkali metal and less iron, although both are nominal charge balanced and possess $Fe^{2+}$ ions. More importantly, the superconducting samples have different iron distribution across the two sites, with almost all occupying the general *16i* site in preference to the *4d* site, which will create crucial differences on the resulting electronic structure.

In summary, we determine the crystal structure of two superconducting analogues of new $A_xFe_{1-x/2}Se_2$ (A = K, Cs) system. They possess I4/m symmetry with a ($\sqrt{5}$ x $\sqrt{5}$ x 1) cell compared with the related $ThCr_2Si_2$ structure, which is created through ordered vacancies of the Fe within the Fe – Se planes, yielding the first examples of a high temperature superconductor with structural holes within the superconducting layers. The iron atoms have a strong tendency to sit on the general *16i* positions leaving the other *4d* sites almost completely empty, in contrast to similar composition that do no possess superconductivity. The control of the both the alkali metal and iron compositions within the structure will be crucial in establishing a phase diagram that maps the superconductivity and antiferromagnetism.

* wbao@ruc.edu.cn, mark.green@nist.gov

**Acknowledgements**
Work at RUC was supported by the National Basic Research Program of China (973 Program) under Grant No. 2011CBA00112, 2010CB923000, 2009CB009100 and 2007CB925001, and by the National Science Foundation of China under Grant No. 11034012, 10834013, 10874244 and 10974254. Work at CUST was supported by the National Science Foundation of China, the Ministry of Science and Technology of China, and the Chinese Academy of Sciences.

**References**
[1] J. Guo, S. Jin, G. Wang, S. Wang, K. Zhu, T. Zhou, M. He, and X. Chen, Physical Review B **82**, 180520 (2010).
[2] F. C. Hsu, et al., Proc. Natl. Acad. Sci. **105**, 14262 (2008).
[3] Y. Kamihara, T. Watanabe, M. Hirano, and H. Hosono, J. Am. Chem. Soc. **130**, 3296 (2008).
[4] G. F. Chen, Z. Li, D. Wu, G. Li, W. Z. Hu, J. Dong, P. Zheng, J. L. Luo, and N. L. Wang, Physical Review Letters **100**, 247002 (2008).
[5] Z. A. Ren, et al., Epl **82**, 57002 (2008).
[6] Z. A. Ren, et al., Chinese Physics Letters **25**, 2215 (2008).
[7] Z. A. Ren, J. Yang, W. Lu, W. Yi, G. C. Che, X. L. Dong, L. L. Sun, and Z. X. Zhao, Materials Research Innovations **12**, 105 (2008).
[8] X. H. Chen, T. Wu, G. Wu, R. H. Liu, H. Chen, and D. F. Fang, Nature **453**, 761 (2008).
[9] R. H. Liu, et al., Physical Review Letters **101**, 087001 (2008).
[10] M. Rotter, M. Tegel, and D. Johrendt, Physical Review Letters **101**, 107006 (2008).
[11] G. F. Chen, et al., Chinese Physics Letters **25**, 3403 (2008).
[12] Y. Qiu, et al., Phys. Rev. Lett. **101**, 257002 (2008).
[13] M. J. Pitcher, D. R. Parker, P. Adamson, S. J. C. Herkelrath, A. T. Boothroyd, R. M. Ibberson, M. Brunelli, and S. J. Clarke, Chemical Communications, 5918 (2008).
[14] X. C. Wang, Q. Q. Liu, Y. X. Lv, W. B. Gao, L. X. Yang, R. C. Yu, F. Y. Li, and C. Q. Jin, Solid State Communications **148**, 538 (2008).
[15] T. M. McQueen, et al., Phys. Rev. B **79**, 014522 (2009).
[16] W. Bao, et al., Physical Review Letters **102**, 247001 (2009).
[17] K. W. Yeh, et al., EPL **84**, 37002 (2008).
[18] M. H. Fang, H. M. Pham, B. Qian, T. J. Liu, E. K. Vehstedt, Y. Liu, L. Spinu, and Z. Q. Mao, Physical Review B **78**, 224503 (2008).
[19] Y. Mizuguchi, F. Tomioka, S. Tsuda, T. Yamaguchi, and Y. Takano, App. Phys. Lett. **94**, 012503 (2009).
[20] B. C. Sales, A. S. Sefat, M. A. McGuire, R. Y. Jin, D. Mandrus, and Y. Mozharivskyj, Physical Review B **79**, 094521 (2009).
[21] P. Zajdel, P.-Y. Hsieh, E. E. Rodriguez, N. P. Butch, J. D. Magil, J. Paglione, P. Zavalij, M. R. Suchomel, and M. A. Green, J. Am. Chem. Soc. **132**, 13000 (2010).
[22] E. E. Rodriguez, P. Zavalij, P. Y. Hsieh, and M. A. Green, J. Am. Chem. Soc. **132**, 10006 (2010).
[23] E. E. Rodriguez, C. Stock, P.-Y. Hsieh, N. Butch, J. Paglione, and M. A. Green, arXiv 1012.0590 (2010).
[24] K. O. Klepp and H. Boller, Monatshefte fuer Chemie **109**, 1049 (1978).
[25] G. Brun, B. Gardes, J. Tedenac, A. Raymond, and M. Maurin, Mater Res Bull **14**, 743 (1979).
[26] R. Berger and C. Van Bruggen, J Less-Common Met **99**, 113 (1984).
[27] L. Haggstrom, H. Verma, S. Bjarman, R. Wappling, and R. Berger, Journal of Solid State Chemistry **63**, 401 (1986).
[28] M. Fang, H. Wang, C. Dong, C. Feng, J. Chen, and H. Q. Yuan, arXiv 1012.5236 (2010).
[29] B. Shen, F. Han, X. Zhu, and H.-H. Wen, arXiv 1012.5637 (2010).
[30] Y. Mizuguchi, H. Takeya, Y. Kawasaki, T. Ozaki, S. Tsuda, T. Yamaguchi, and Y. Takano, arXiv 1012.4950 (2010).
[31] D. M. Wang, J. B. He, T.-L. Xia, and G. F. Chen, arXiv 1101.0789 (2011).
[32] A. F. Wang, et al., arXiv 1012.5525 (2010).
[33] A. Krzton-Maziopa, Z. Shermadini, E. Pomjakushina, V. Pomjakushin, M. Bendele, A. Amato, R. Khasanov, H. Luetkens, and K. Conder, arXiv 1012.3637 (2010).
[34] J. J. Ying, et al., arXiv 1101.1234 (2011).
[35] Z. Wang, et al., arXiv 1101.2059 (2011).
[36] W. Bao, Q. Huang, G. F. Chen, M. A. Green, D. M. Wang, J. B. He, X. Q. Wang, and Y. Qiu, arXiv 1102.0830 (2011).
[37] J. Bacsa, A. Y. Ganin, Y. Takabayashi, K. E. Christensen, K. Prassides, M. J. Rosseinsky, and J. B. Claridge, arXiv 1102.0488 (2011).
[38] J. J. Ying, et al., arXiv 1012.5552 (2010).
[39] W. Yu, L. Ma, J. B. He, D. M. Wang, T.-L. Xia, and G. F. Chen, arXiv 1101.1017 (2011).
[40] J. Zhao, et al., Nature Materials **7**, 953 (2008).
[41] A. Kreyssig, et al., Physical Review B **78**, 184517 (2008).
[42] C. H. Lee, et al., Journal of the Physical Society of Japan **77**, 083704 (2008).